\documentclass[12pt]{article}
\usepackage{epsfig}
\usepackage{graphicx}
\usepackage{longtable}
\usepackage{lscape}
\usepackage{rotating}

\begin{document}

\noindent  {\Large \bf  Radiopure tungstate and molybdate crystal
scintillators for double beta decay experiments}

\vskip 1.cm

\noindent {\bf F.A.~Danevich}

\vskip 0.3cm

\noindent {\it Institute for Nuclear Research, 03028 Kyiv,
Ukraine}

\noindent {\it CSNSM, Univ. Paris-Sud, CNRS/IN2P3, Universit\'{e}
Paris-Saclay, 91405 Orsay, France}

\vskip 0.8cm

\noindent Crystal scintillators are very promising detectors to
investigate double beta decay of atomic nuclei. Recent
achievements in development and application of tungstate and
molybdate crystal scintillators as well as prospects for the next
generation double beta decay experiments are discussed.

\vskip 0.4cm

\noindent {\bf Keywords:} Double beta decay; Tungstate and
Molybdate crystal scintillators; Low counting experiment

\section{Introduction}

Double beta ($2\beta$) decay of atomic nuclei is the rarest
nuclear process observed by human being. Despite the two neutrino
mode ($2\nu$) of the decay is allowed by all the known laws, its
probability is extremely low. After the great efforts of
experimentalists over seventy years, by utilizing different
methods: geochemical, radiochemical, direct counting, the decay is
observed in eleven nuclei with the half-lives in the range
$T_{1/2}\sim 10^{18}-10^{24}$ yr
\cite{Tretyak:2002,Saakyan:2013,Barabash:2015}.

The neutrinoless mode of double beta decay ($0\nu2\beta$) is
forbidden in the Standard Model of particle physics since the
decay breaks the lepton number and calls for the neutrinos to be
Majorana particles with nonzero mass \cite{Schechter:1982}. The
process is considered as one of the most powerful tools to study
properties of neutrino and weak interaction, and to test the
Standard Model of particle physics
\cite{Barrea:2012,Rodejohann:2012,Deppisch:2012,Bilenky:2015,Delloro:2016,Vergados:2016}.

In contrast to the $2\nu2\beta^-$ decay that is already observed,
even the most sensitive experiments give only half-life limits at
level of $T_{1/2}> 10^{24}-10^{26}$ yr on the $0\nu$ mode of decay
(we refer reader to the reviews
\cite{Elliott:2012,Giuliani:2012,Cremonesi:2014,Gomes:2015,Sarazin:2015,
Delloro:2016} and to the recent experimental searches
\cite{GERDA,EXO-200,CUORE,NEMO-3,Gando:2017}). The experiments set
limits on the effective Majorana neutrino mass at the level of
$\langle m_{\nu} \rangle \sim$ 0.1 -- 1 eV. The range is due to
the ambiguity of the theoretical calculations of nuclear matrix
elements of $0\nu2\beta^-$ decay \cite{Engel:2017}.

The sensitivity of the experiments to search for the double
electron capture ($2\varepsilon$), electron capture with positron
emission ($\varepsilon\beta^+$), and double positron emission
($2\beta^+$) is much lower. In contrast to the $2\beta^-$ decay,
even the allowed two neutrino mode of these processes has not yet
been detected clearly (see, e.g, reviews
\cite{Tretyak:2002,Maalampi:2013}). However, development of
experimental techniques to search for the processes is requested,
taking into account a capability to refine the possible
contribution of the right-handed currents to the $0\nu2\beta^-$
decay rate if the decay will be observed \cite{Hirsch:1994}. In
addition, a resonant enhancement of the $0\nu2\varepsilon$ process
can increase the decay rate by several orders of magnitude in the
case of coincidence between the released energy and the energy of
an excited state of the daughter nucleus
\cite{Winter:1955,Voloshin:1982,Bernabeu:1983,Sujkowski:2004,Krivoruchenko:2011,Kotila:2014,Blaum:2017}.

The goal of the next generation $2\beta^-$ experiments is to
explore the inverted hierarchy of the neutrino mass (the effective
Majorana mass of neutrino $\langle m_{\nu}\rangle\approx
0.02-0.05$ eV). Such a sensitivity can be reached in experiments
able to detect an extremely low level of $0\nu2\beta^-$ activity
with the half-lives in the range $T_{1/2} \sim 10^{26} - 10^{27}$
yr. The realization of such ambitious plans requires construction
of detectors containing a large number of $2\beta^-$ active nuclei
($10^{27}-10^{28}$ nuclei, $\sim 10^3-10^4$ moles of isotope of
interest), extremely low (ideally zero) radioactive background,
high detection efficiency and ability to distinguish the
$0\nu2\beta^-$ decay events, in particular, as high as possible
energy resolution. It should be stressed that a low energy
resolution is acceptable as far as an experiment set a limit on
the process searched for. However, in a case of an effect evidence
the poor energy resolution become to be a controversial factor in
the effect interpretation.

Taking into account an extremely low decay probability and the
ambiguity of the theoretical calculations of the nuclear matrix
elements, the experimental program should include several nuclei
candidates. First of all, confirmation in different experiments
will be naturally requested in a case of a positive evidence of
the $0\nu2\beta^-$ decay in one nucleus. Moreover, data on
$0\nu2\beta^-$ decay rate in several nuclei will be useful to
adjust theoretical approaches to calculate the nuclear matrix
elements \cite{Bilenky:2002}.

Therefore, choice of candidate nuclei is determined by the scale
of experiments (several hundred kilograms of isotope of interest),
extreme requirements to background and possibility of calorimetric
experiments ("source = detector") to provide high detection
efficiency and energy resolution. In this paper we would like to
argue that scintillation detectors, especially low temperature
scintillating bolometers, are very promising to realize large
scale high sensitivity $0\nu2\beta^-$ decay experiments with
several nuclei.

\section{Scintillation detectors in double beta decay experiments}
\label{sec:sc2b}

Crystal scintillators possess certain advantages for double beta
decay experiments thanks to presence of the element of interest in
the crystal scintillator compound, that provides a high detection
efficiency for the effect searched for. Cadmium and zinc
tungstates, zinc selenide crystal scintillators were studied as
detector candidates for double beta experiments in
\cite{Danevich:1989a}, lead molybdate was proposed to search for
$2\beta^-$ decay of $^{100}$Mo in \cite{Minova:1992}.

The most sensitive $2\beta$ experiments with crystal scintillators
are reported in Table \ref{tab1}. We would like to repeat that the
current sensitivity of the experiments to search for
$2\varepsilon$, $\varepsilon\beta^+$ and $2\beta^+$ decays is a
few orders of magnitude lower than that of the $2\beta^-$
experiments. Therefore, the listen limits for the $2\varepsilon$
and $\varepsilon\beta^+$ decays correspond broadly to the level of
the most sensitive investigations in the field. It should be
stressed that the half-life values and limits for the $2\beta^-$
decay of $^{100}$Mo are very preliminary results of the R\&Ds in
progress aiming at construction of large scale cryogenic
experiments to search for $0\nu2\beta^-$ decay in $^{100}$Mo with
a sensitivity a few orders of magnitude higher than that the
reported in Table \ref{tab1} (the projects will be discussed in
Section \ref{sec:prosp}).

\noindent
\begin{table}[ht]
\caption{The most sensitive $2\beta$ experiments with crystal
scintillators.}
\begin{center}
\begin{tabular}{|l|l|l|l|}
 \hline
 $2\beta$ transition                & Scintillator          & Main results: half-life (channels)                                                        & Ref. \\
  ~                                 & ~                     & ~                                                                                         & ~         \\
  \hline
 $^{40}$Ca$\rightarrow$$^{40}$Ar    & CaWO$_4$              & $\geq 9.9\times10^{21}$ yr ($2\nu2K$)                                                     & \cite{Angloher:2016} \\
 ~                                  & ~                     & $\geq 1.4\times10^{22}$ yr ($0\nu2\varepsilon$)                                                     & \cite{Angloher:2016} \\
  \hline
 $^{48}$Ca$\rightarrow$$^{48}$Ti    & CaF$_2$(Eu)           & $\geq 5.8\times10^{22}$ yr ($0\nu2\beta^-$)                                                 & \cite{Umehara:2008} \\
  \hline
 $^{64}$Zn$\rightarrow$$^{64}$Ni    & ZnWO$_4$              & $\geq 1.1\times10^{19}$ yr ($2\nu2K$)                                                     & \cite{Belli:2011a} \\
  ~                                 & ~                     & $\geq 9.4\times10^{20}$ yr ($2\nu\varepsilon\beta^+$)                                     & \cite{Belli:2011a} \\
  \hline
 $^{100}$Mo$\rightarrow$$^{100}$Ru  & ZnMoO$_4$             & $=[7.15 \pm 0.37(stat.) \pm 0.66(syst.)]\times 10^{18}$ yr ($2\nu2\beta^-$)   & \cite{Cardani:2014} \\
 ~                                  & Li$_2^{100}$MoO$_4$   & $=[6.92 \pm 0.06(stat.)\pm 0.36(syst.)]\times 10^{18}$ yr ($2\nu2\beta^-$)    & \cite{Poda:2017} \\
 ~                                  & Zn$^{100}$MoO$_4$     & $\geq 2.6\times10^{22}$ yr ($0\nu2\beta^-$)                                                 & \cite{Poda:2017} \\
 ~                                  & $^{48depl}$Ca$^{100}$MoO$_4$ & $\geq 4.0\times10^{21}$ yr ($0\nu2\beta^-$)                                          & \cite{So:2012a} \\
 \hline
 $^{106}$Cd$\rightarrow$$^{106}$Pd  & $^{106}$CdWO$_4$      & $\geq 1.1\times10^{21}$ yr ($2\nu\varepsilon\beta^+$)                                     & \cite{Belli:2016a} \\
 ~                                  & ~                     & $\geq 2.2\times10^{21}$ yr ($0\nu\varepsilon\beta^+$)                                     & \cite{Belli:2012} \\
 \hline
 $^{116}$Cd$\rightarrow$$^{116}$Sn  & $^{116}$CdWO$_4$      & $=[2.69 \pm 0.02(stat.)\pm 0.14(syst.)]\times 10^{19}$ yr ($2\nu2\beta^-$)    & \cite{Polischuk:2017} \\
 ~                                  & ~                     & $\geq 2.4\times10^{23}$ yr ($0\nu2\beta^-$)                                                 & \cite{Polischuk:2017} \\
 \hline
 $^{130}$Ba$\rightarrow$$^{130}$Xe  & BaF$_2$               & $\geq 2.0\times10^{17}$ yr ($0\nu\varepsilon\beta^+$)                                     & \cite{Cerulli:2004} \\
 \hline
 $^{136}$Ce$\rightarrow$$^{136}$Ba  & CeCl$_3$               & $\geq 3.2\times10^{16}$ yr ($2\nu2K$)                                                    & \cite{Belli:2011b} \\
 \hline
 $^{160}$Gd$\rightarrow$$^{160}$Dy  & Gd$_2$SiO$_5$(Ce)     & $\geq 1.3\times10^{21}$ yr ($0\nu2\beta^-$)                                                 & \cite{Danevich:2001} \\
 \hline
 $^{180}$W$\rightarrow$$^{180}$Hf   & CaWO$_4$              & $\geq 3.1\times10^{19}$ yr ($2\nu2K$)                                                     & \cite{Angloher:2016} \\
 ~                                  & ~                     & $\geq 9.4\times10^{18}$ yr ($2\nu2\varepsilon$)                                           & \cite{Angloher:2016} \\
 \hline
\end{tabular}
\end{center}
\label{tab1}
\end{table}

Characteristics of the $2\beta$ isotopes present in the already
developed tungstate and molybdate crystal scintillators are listed
in Table \ref{tab2}.

\clearpage
\begin{table}[ht]
\caption{Characteristics of $2\beta$ isotopes present in tungstate
and molybdate crystal scintillators.}
\begin{center}
\begin{tabular}{|l|l|l|l|}
  \hline
 $2\beta$ transition                & $Q_{2\beta}$ (keV) \cite{Wang:2017}   & Isotopic abundance (\%) \cite{Meija:2016}  & Decay channel \\
 \hline
 $^{40}$Ca$\rightarrow$$^{40}$Ar    & 193.51(2)             &   96.941(156) & $2\varepsilon$  \\
 \hline
 $^{46}$Ca$\rightarrow$$^{46}$Ti    & 988.4(2.2)            &   0.004(3)    & $2\beta^-$  \\
 \hline
 $^{48}$Ca$\rightarrow$$^{48}$Ti    & 4268.08(8)            &   0.187(21)   & $2\beta^-$  \\
 \hline
$^{64}$Zn$\rightarrow$$^{64}$Ni     & 1094.9(7)             &   49.17(75)   & $2\varepsilon$, $\varepsilon\beta^+$  \\
 \hline
$^{70}$Zn$\rightarrow$$^{70}$Ge     & 997.1(2.1)            &   0.61(10)    & $2\beta^-$  \\
 \hline
$^{84}$Sr$\rightarrow$$^{84}$Kr     & 1789.8(1.2)           &   0.56(2)     & $2\varepsilon$, $\varepsilon\beta^+$  \\
 \hline
$^{92}$Mo$\rightarrow$$^{92}$Zr     & 1650.45(19)           &  14.649(106)  & $2\varepsilon$, $\varepsilon\beta^+$  \\
 \hline
$^{98}$Mo$\rightarrow$$^{98}$Ru     &  109(6)               & 24.292(80)    & $2\beta^-$  \\
 \hline
$^{100}$Mo$\rightarrow$$^{100}$Ru   &  3034.36(17)          & 9.744(65)     & $2\beta^-$  \\
 \hline
 $^{106}$Cd$\rightarrow$$^{106}$Pd  & 2775.39(10)           & 1.245(22)     & $2\varepsilon$, $\varepsilon\beta^+$, $2\beta^+$  \\
 \hline
$^{108}$Cd$\rightarrow$$^{108}$Pd   & 271.8(8)              & 0.888(11)     & $2\varepsilon$  \\
 \hline
$^{114}$Cd$\rightarrow$$^{114}$Sn   & 544.79(28)            &  28.754(81)   & $2\beta^-$  \\
 \hline
$^{116}$Cd$\rightarrow$$^{116}$Sn   & 2813.49(13)           &  7.512(54)    & $2\beta^-$  \\
 \hline
$^{180}$W$\rightarrow$$^{180}$Hf    &  143.23(28)          &  0.12(1)       & $2\varepsilon$  \\
 \hline
$^{186}$W$\rightarrow$$^{186}$Os    &  491.4(1.2)          &  28.43(19)     & $2\beta^-$  \\

 \hline
\end{tabular}
\end{center}
\label{tab2}
\end{table}

\section{Requirements to crystal scintillators}
\label{sec:req}

Sensitivity of a $0\nu2\beta^-$ counting experiment can be
expressed by the following formula \cite{Zdesenko:2002}:

\begin{equation}
\lim T_{1/2} \sim \varepsilon \cdot \delta \sqrt{\frac{m \cdot
t}{R \cdot BG}},
 \label{eq1}
\end{equation}

\noindent where $\varepsilon$ is detection efficiency of the
$0\nu2\beta^-$ effect, $\delta$ is the concentration of the
isotope of interest, $t$ is the measurement time, $m$ is the mass,
$R$ is the energy resolution and $BG$ is the background per unit
mass, time and energy. In a case of very low background ("zero
background experiment") the sensitivity is proportional to the
exposure of experiment $m \cdot t$:

\begin{equation}
 \lim T_{1/2} \sim \varepsilon \cdot \delta \frac{m \cdot t}{\lim S},
 \label{eq2}
\end{equation}

\noindent where $\lim S$ is number of events that can be excluded
at a given confidence level. E.g., according to the Feldman and
Cousins procedure \cite{Feldman:1998} $\lim S = 2.44, 3.28$ and
$3.94$ at 90\% confidence level (C.L.) in the case of detected
$0$, $1$ and $2$ events, respectively.

From the formulas \ref{eq1} and  \ref{eq2} one can immediately see
the advantage of scintillation $0\nu2\beta^-$ counting experiment:
a high detection efficiency $\varepsilon$ (approaching to
$\approx100\%$ with increase of the crystal scintillator volume
and density) thanks to presence of the element of interest.
$\delta$ cannot be 100\% in a crystal scintillator (as, e.g., in
the high purity germanium detectors used in the $^{76}$Ge
experiments). Nevertheless, typically $\delta$ is rather high in
tungstate and molybdate crystals. The concentration can be
increased by production of crystal scintillators from isotopically
enriched materials (see Section \ref{sec:iso}).

The main properties of tungstate and molybdate crystal
scintillators that are interesting for $2\beta$ experiments are
reported in Table \ref{tab3}.

\clearpage
 \begin{landscape}
 \begin{center}
 \begin{longtable}{|l|l|l|l|l|l|l|l|l|l|}
 \caption{Properties of tungstate and molybdate crystal
scintillators. Wavelength of scintillation emission maximum is
denoted as $\lambda_{max}$; FWHM is full width at half of maximum;
$FOM$ is factor of merit of particle discrimination capability
(see formula \ref{eq:sens1}); $FOM$ for room temperature
scintillators is given for $\approx 5$ MeV $\alpha$ particles and
$\approx 1$ MeV $\gamma$ quanta, while for milli-Kelvin
measurements $FOM$ is presented mainly for energy of $\alpha$
particles and $\gamma$ quanta above 2.5 MeV.}\\
  \hline
 Scintillator       & Density   & Melting           & Index of      & $\lambda_{max}$   &  \multicolumn{2}{|c|}{Energy resolution, FWHM} &  \multicolumn{2}{|c|}{Particle discrimination, $FOM$}   & Ref. \\
 \cline{6-7}
 \cline{8-9}
 ~                  &(g/cm$^3$) & point ($^{\circ}$C) & refraction  & (nm)              & Room          & milli-Kelvin,                  & Room          & milli-Kelvin                      & ~ \\
 ~                  & ~         & ~                 & ~             & ~                 & temperature,   & at 2615 keV                   & temperature   & ~                                 & ~ \\
 ~                  & ~         & ~                 & ~             &                   & at 662 keV    & ~                             & ~             & ~                                 & ~ \\
 \hline
 MgWO$_4$           & 5.66      & 1358              & ~             & $470$             & $9.1\%$       & ~                             & 5.2           & ~                                 & \cite{Danevich:2009} \\
 \hline
 CaWO$_4$           & 6.1       & $1570-1650$       & 1.94          & $425$             & $6.4\%$       & 6.7 keV$^{\ast\ast}$          & 5.9           & ~                                 & \cite{Zdesenko:2005,Cozzini:2004,Danevich:2014} \\
 \hline
 ZnWO$_4$           & 7.8       & 1200              & $2.1 - 2.2$   & 480               & $8.5\%$       & 4.7 keV$^{\ast\ast}$          & 5.2           & ~                                 & \cite{Danevich:2005,Nagornaya:2008,Casali:2016} \\
 \hline
 CdWO$_4$           & 7.9       & 1325              & $2.2 - 2.3$   & 480               & $6.8\%$       & 6.3 keV                       & 5.8           & 15                                & \cite{Bardelli:2006,Arnaboldi:2010a,Artusa:2014} \\
 \hline
 PbWO$_4$           & 8.28      & 1123              & 2.2           & $450$             &23\%$^{\ast}$ & 15.7 keV                       & 2.4$^{\ast}$  & ~                                 & \cite{Danevich:2006,Bardelli:2008,Beeman:2013} \\
 \hline
 Li$_2$MoO$_4$      & 3.05      & $701\pm2$         & 1.44          & $590$             & ~             & $4-6$ keV                     & ~             & $9 - 18$                        & \cite{Bekker:2016,Armengaud:2017} \\
 \hline
 CaMoO$_4$          & $4.3$ &   $1445-1480$         & 2.0           & 520               & 10.3\%        & 10.9 keV                      & 2.2           & 7.6                               & \cite{Annenkov:2008,Kim:2015} \\
 \hline
 ZnMoO$_4$          & 4.3       & $1003\pm5$        & $1.89-1.96$   & $625$             &               & $4-22$ keV                    & ~             & $8 - 21$                        & \cite{Armengaud:2017,Chernyak:2013,Beeman:2012} \\
 \hline
 SrMoO$_4$           & 4.67     & 1380              & ~             & 506               &               & ~                             & ~             & ~                                 & \cite{Jiang:2013} \\
 \hline
 PbMoO$_4$           & 6.95     & 1060              & 2.4           & 540               &               & ~                             & 2$^{\ast\ast\ast}$ & ~                            & \cite{Danevich:2010} \\
\hline

 \multicolumn{8}{l}{$^{\ast}$~At temperature $-25^{\circ}$C.} \\
 \multicolumn{8}{l}{$^{\ast\ast}$~For 2.31 MeV $\alpha$.} \\
 \multicolumn{8}{l}{$^{\ast\ast\ast}$~At temperature $-196^{\circ}$C.} \\

\end{longtable}
\end{center}
\label{tab3}
\end{landscape}

Energy resolution of the convenient scintillation detectors at
room temperature is at the level of several \%. Around the
tungstate crystals discussed as $2\beta$ detectors, the highest
energy resolution (full width at half of maximum, FWHM, for
$\gamma$ quanta of $^{137}$Cs with energy 662 keV) was achieved
with CaWO$_4$ ($6.4\%$, \cite{Danevich:2014}), CdWO$_4$ ($6.8\%$,
\cite{Bardelli:2006}), ZnWO$_4$ ($8.5\%$, \cite{Nagornaya:2008})
and MgWO$_4$ ($9.1\%$, \cite{Danevich:2009}). As for molybdates,
only CaMoO$_4$ shows reasonable scintillation efficiency and
energy resolution FWHM$~=10.3\%$ \cite{Annenkov:2008}.

The energy resolution obtained with the scintillators at low
temperatures (denoted in the Table  \ref{tab3} as milli-Kelvin
temperatures) is significantly higher, that is a great advantage
of the bolometric detection technique.

Most of scintillators posses particle-discrimination capability
that allows a substantial suppression of background caused by the
internal and surface radioactive contamination of the crystal
scintillators by $\alpha$ active nuclides, particularly by
uranium, radium and thorium. The particle discrimination
capability (denoted here as $FOM$, factor of merit) can be
determined by the following formula:

\begin{equation}
FOM = |SI_{\alpha}-SI_{\gamma}|/
\sqrt{\sigma^2_{\alpha}+\sigma^2_{\gamma}},
 \label{eq:sens1}
\end{equation}

\noindent where $SI_{\alpha}$ ($\sigma_{\alpha}$) and
$SI_{\gamma}$ ($\sigma_{\gamma}$) are mean values (standard
deviations) of Shape Indicator parameters for $\alpha$ particles
and $\gamma$ quanta distributions. The $SI$ parameters can be
estimated, e.g., by using the optimal filter method (see the
application of the method to CdWO$_4$ scintillators in
\cite{Fazzini:1998,Bardelli:2006}), or by calculation of the mean
time of scintillation signal (see, e.g., \cite{Bardelli:2008}).

As one can see in Table \ref{tab3}, cryogenic scintillating
bolometers possess a much higher particle discrimination
capability than that the conventional scintillation detectors. The
discrimination can be realized by analysis of heat and
scintillation signals from a crystal scintillator (taking into
account a much lower scintillation yield for ions\footnote{Zinc
selenide scintillators at milli-Kelvin temperature show an
opposite property: a higher emission from $\alpha$ particles than
that from $\beta$ particles and $\gamma$
quanta\cite{Arnaboldi:2011}.}, in particular for $\alpha$
particles \cite{Tretyak:2010}) or by only thermal pulse profile
analysis \cite{Gironi:2010}.

The most important source of scintillation detector background,
that finally limits a $2\beta$ experiment sensitivity, is
radioactive contamination of crystal scintillators. The issue is
discussed in the following section.

\subsection{Radiopurity}
\label{sec:rp}

Radioactive contamination of a scintillation material plays a key
role in $2\beta$ experiments that require as low as possible
(ideally zero) background counting rate in the region of interest
(ROI). Some data on radioactive contamination of tungstate and
molybdate crystal scintillators are presented in Table \ref{tab4}.

\clearpage
 \begin{landscape}
 \begin{center}
 \begin{longtable}{|l|l|l|l|l|l|l|}
\caption{Radioactive contamination of tungstate and molybdate
crystal scintillators. Radioactive isotopes of tungstate
($^{180}$W, $\alpha$ active) and molybdate ($^{100}$Mo, $2\beta^-$ active) scintillators are not specified.
The lowest reported limit is presented if no activity of the nuclide was detected. The lowest limit and the biggest value of activity (or the range of values) are given if several samples of the scintillator were studied.} \\
  \hline
 Scintillator       &\multicolumn{5}{|c|} {Radioactive contamination (mBq/kg)}                                                  & Reference \\
 \cline{2-6}
  ~                 & $^{40}$K          & $^{226}$Ra    & $^{228}$Th    & total $\alpha$            & Inherent radionuclides             &  ~ \\
  ~                 &  ~                & ~             & ~             & ~                         & in scintillator compound           & ~\\
%  ~                 &  ~                & ~             & ~             & ~                         &                   & ~\\
 \hline
 MgWO$_4$           & $<1.6\times 10^3$ & $<50$         & $<50$         & $5.7(4)\times 10^3$    &    ~                      & \cite{Danevich:2009} \\
 \hline
 CaWO$_4$           & $\leq 12$         & $6-7$         & $<0.2-0.6(2)$ & $400-930$                 & $^{48}$Ca ($2\beta^-$)      & \cite{Zdesenko:2005,Cebrian:2004} \\
 \hline
 ZnWO$_4$           & $<0.02$           & $0.002(1)-0.025(6)$ & $0.002(2)-0.018(2)$ & $0.18(3)-2.3(2)$                & ~                                             & \cite{Belli:2011c} \\
 \hline
 CdWO$_4$           & $<1.7-3.6(2)$     & $<0.007$      & $<0.003-0.008(4)$  & 0.26(4)              & $^{113}$Cd ($\beta$), $^{113m}$Cd ($\beta$), $^{116}$Cd ($2\beta^-$) & \cite{Danevich:1996,Belli:2007,Georgadze:1996} \\
 \hline
 $^{106}$CdWO$_4$   & $<1.4$            & 0.012(3)      & 0.042(4)      & 2.1(2)                    & $^{113m}$Cd ($\beta$), $^{113}$Cd ($\beta$), $^{116}$Cd ($2\beta^-$) & \cite{Belli:2012} \\
 \hline
 $^{116}$CdWO$_4$   & $\leq 0.9-0.3(1)$ & $\leq 0.004$  & $0.010(3)-0.062(6)$   & $1.4(1)-2.93(2)$ & $^{113m}$Cd ($\beta$), $^{113}$Cd ($\beta$), $^{116}$Cd ($2\beta^-$) & \cite{Danevich:2003a,Danevich:2003b,Barabash:2011} \\
 ~   & ~ & ~ & ~   & ~ & ~ & \cite{Poda:2013,Barabash:2016a} \\
 \hline
 PbWO$_4$           & ~                 & $1.40(4)$     & $0.051(8)$$^{\ast}$   & ~                 &$^{210}$Pb and daughters ($\beta$, $\alpha$) & \cite{Beeman:2013} \\
 \hline
 Li$_2$MoO$_4$      & $\leq 3.2-62(2)$  & $\leq 0.028-0.13(2)$  & $\leq 0.018$  & ~                 & ~                          & \cite{Armengaud:2017} \\
 \hline
 Li$_2^{100}$MoO$_4$ & $\leq 3.5$       & $\leq 0.007$          & $\leq 0.006$  & ~                 & ~                          & \cite{Armengaud:2017} \\
 \hline
 ZnMoO$_4$          & ~                 & $\leq 0.006$          & $\leq 0.005$  & ~                 & ~                         & \cite{Armengaud:2017} \\
 \hline
 Zn$^{100}$MoO$_4$   & ~                 & $0.014(3)-0.023(4)$      & $\leq 0.008$          & ~                 & ~                          & \cite{Armengaud:2017} \\
 \hline
 CaMoO$_4$          & $<1$              & $0.13(4)-2.5(5)$          & $0.04(2)-0.42(17)$    & ~                 &  $^{48}$Ca ($2\beta^-$)   & \cite{Annenkov:2008} \\
 \hline
 $^{48depl}$Ca$^{100}$MoO$_4$ & ~  & $0.065$        &  $<0.05$              & ~                 & $^{48}$Ca ($2\beta^-$)   & \cite{Luqman:2017} \\

 \hline

 \multicolumn{7}{l}{$^{\ast}$~Activity of $^{232}$Th.} \\

 \label{tab4}
 \end{longtable}
 \end{center}
 \end{landscape}

The main radioactive contaminations of scintillation materials are
naturally occurring radioactive elements, as uranium (consists of
isotopes $^{235}$U and $^{238}$U), thorium ($^{232}$Th) with their
daughters, and potassium ($^{40}$K). The secular equilibrium of
the U/Th chains is typically broken in scintillation materials
that pass through chemical and physical procedures of
purification, synthesis and crystal growth. It means that the
activities of $^{238}$U, $^{230}$Th, $^{226}$Ra, $^{210}$Pb, and
$^{210}$Po in the $^{238}$U family should be considered
separately. Similarly, activities of $^{232}$Th, $^{228}$Ra and
$^{228}$Th from the $^{232}$Th family, and activities of
$^{235}$U, $^{231}$Pa and $^{227}$Ac from the $^{235}$U family can
be different in a sample of scintillator. For instance, the
equilibrium of $^{238}$U chain is strongly broken in CaWO$_4$
\cite{Zdesenko:2005}.

Some scintillators contain valuable impurities of primordial
radioactive isotopes of rare-earth elements. As an example we
again refer to CaWO$_4$ where $\alpha$ radioactive $^{144}$Nd,
$^{147}$Sm, and $^{152}$Gd were detected at a mBq/kg level
\cite{Zdesenko:2005,Cozzini:2004}. Anthropogenic
$^{90}$Sr~$-^{90}$Y and $^{137}$Cs can also be present in
scintillators, especially after the Chernobyl (in 1986), and the
Fukushima Daiichi (2012) nuclear disasters\footnote{E.g., it was
quite expectable to find $^{134}$Cs and $^{137}$Cs in CsI(Tl)
scintillators \cite{Lee:2007}.}.

Radioactive contamination of crystal scintillators is determined
substantially by its chemical formula and possibilities of initial
materials purification. For instance CdWO$_4$ and ZnWO$_4$
crystals have typically rather low level of internal
contamination, while CaWO$_4$ crystals are rather contaminated,
mainly by radium (see Table \ref{tab4}). Same dependence of radium
activity on presence of calcium is observed in ZnMoO$_4$,
Li$_2$MoO$_4$ (highly radiopure materials with only limits on
$^{226}$Ra and $^{228}$Th activities on the level of $<0.005$
mBq/kg) versus CaMoO$_4$ crystals, where activity of $^{226}$Ra is
still on the level of $\sim0.1$ mBq/kg, despite the rather strong
efforts to reduce the contamination. The contamination of calcium
containing crystals is due to the chemical similarity of calcium
and radium, and due to the certain difficulties to purify calcium
with effective physical purification methods, like vacuum
distillation.

The main source of scintillation material radioactive
contamination is radioactive contamination of the materials used
for powder synthesis and then for the crystal growth. Possible
effect of ceramic details contamination of the growing apparats on
the ZnWO$_4$ crystal scintillators radiopurity was investigated in
\cite{Belli:2011c}. There is no indication of tungstate and
molybdate crystal scintillators pollution by the growing process
(such as platinum contamination of TeO$_2$ crystals grown in
platinum crucibles observed by the CUORE collaboration
\cite{Arnaboldi:2010b}). Nevertheless, possible effects of the
radioactive contamination of the components of the growing set-ups
on the radiopurity of the crystal scintillators should be
addressed in the future developments of highly radiopure crystal
scintillators.

Recrystallization can improve significantly the radiopurity level
of tungstate and molybdate crystals
\cite{Danevich:2011,Armengaud:2017}. Recently the concentration of
thorium was reduced by one order of magnitude, down to 0.01 mBq/kg
($^{228}$Th), by recrystallization of the $^{116}$CdWO$_4$ sample
No. 3 (Fig. 1 in \cite{Barabash:2011}) by the low-thermal-gradient
Czochralski method. The total $\alpha$ activity of uranium and
thorium daughters was reduced by a factor $\approx3$, down to 1.6
mBq/kg \cite{Barabash:2016a}.

Cosmogenic radionuclides, i.e. created by high energy neutrons
(mainly of cosmic rays origin), were observed in tungstate
scintillators: $^{65}$Zn in ZnWO$_4$ \cite{Belli:2011c}, and
$^{110m}$Ag in CdWO$_4$ \cite{Poda:2013} crystal scintillators on
a $\sim0.1$ mBq/kg level. To reduce cosmogenic activation the
production line and transportation of the ready-to-use
scintillation elements to the experiment site should be organized
in a way providing minimal cosmogenic activation.

Some scintillators have radioactive elements in their chemical
formula. Cadmium tungstate scintillation crystals contain
primordial $\beta$ radioactive isotope $^{113}$Cd
($Q_{\beta}=323.83(27)$ keV, isotopic abundance 12.227(7)\%,
$T_{1/2}=8.04(5)\times 10^{15}$ yr). Beta active $^{210}$Pb
($Q_{\beta}=63.5(5)$ keV, $T_{1/2}=22.20(22)$ yr) is usually
present in PbWO$_4$ and PbMoO$_4$ on the level of hundreds Bq/kg.
The $^{210}$Pb activity can be reduced by several orders of
magnitude by production of lead containing scintillators from
archaeological lead
\cite{Alessandrello:1998,Beeman:2013,Belli:2016a}. Despite being a
very weak, the $2\beta^-$ activity of $^{48}$Ca, $^{100}$Mo,
$^{116}$Cd, $\alpha$ activity of $^{180}$W can also be sources of
background in low counting experiments. For instance, presence of
$2\beta^-$ active $^{48}$Ca in CaMoO$_4$ makes problematic
application of the crystal scintillator in high sensitivity
experiments to search for $0\nu2\beta^-$ decay of $^{100}$Mo,
while the $2\nu2\beta^-$ decays of $^{100}$Mo will generate
unavoidable background in the ROI of $^{116}$Cd in a detector
based on CdMoO$_4$ crystal scintillators \cite{Xue:2017}.

\subsection{Isotopically enriched scintillators}
\label{sec:iso}

High concentration of isotope of interest is an important
requirement to scintillation materials to be used in $2\beta$
decay experiments that suggests development of scintillators from
enriched isotopes. The request for radiopure crystal scintillators
from isotopically enriched materials imposes specific requirements
to the production process: maximal yield of the ready to use
scintillation elements and minimal losses of the enriched
material. Typical purity grade of isotopically enriched materials
(both chemical and radioactive contaminations) is much worse than
that required for radiopure crystal growth. Therefore, a
production cycle of enriched scintillators should include
development of specific purification, crystal growth and treatment
protocols, recovery of costly enriched materials from all the
scraps \cite{Danevich:2012}.

First crystal scintillators from enriched materials were developed
by der Mateosian and Goldhaber \cite{Mateosian:1966}.
Europium-activated calcium fluoride crystal scintillators enriched
in $^{48}$Ca ($^{48}$CaF$_2$(Eu)) and depleted in $^{48}$Ca
($^{48depl}$CaF$_2$(Eu)) were grown to search for the $2\beta^-$
decay of $^{48}$Ca at the level of sensitivity $\lim
T_{1/2}^{0\nu2\beta^-}=2\times10^{20}$ yr \cite{Mateosian:1966}.
However, the crystals enriched in $^{48}$Ca were substantially
contaminated by U/Th.

Radiopure cadmium tungstate crystal scintillator enriched in
$^{116}$Cd ($^{116}$CdWO$_4$) was developed for the Solotvina
experiment \cite{Danevich:1989b} by using the convenient
Czochralski growth process. In the experiment at the Solotvina
Underground laboratory the two neutrino mode of the $2\beta^-$
decay of $^{116}$Cd was observed. For the neutrinoless $2\beta^-$
decay of $^{116}$Cd the half-life limit $T_{1/2}\geq 1.7\times
10^{23}$ yr was set at 90\% C.L., which corresponds to the
effective Majorana neutrino mass limit $\langle m_{\nu} \rangle
\leq 1.7$ eV \cite{Danevich:2003b}.

Cadmium tungstate crystal scintillator from enriched to 66\%
isotope cadmium 106 ($^{106}$CdWO$_4$) with mass 231 g was
developed for the double beta experiments with $^{106}$Cd, that is
one of the most promising double beta plus decay nuclei. The deep
purification of initial materials
\cite{Bernabey:2008,Kovtun:2011}, particularly application of
vacuum distillation with filtering on getter filters to purify the
enriched cadmium samples, together with the low-thermal-gradient
Czochralski method to grow the crystal
\cite{Pavlyuk:1993,Galashov:2009}, resulted in unprecedented high
optical quality of the crystal scintillator \cite{Belli:2010}. The
$^{106}$CdWO$_4$ crystal is now used in experiments to search for
$2\beta$ processes in $^{106}$Cd, with preliminary results
reported in \cite{Belli:2012,Belli:2016a}. In the last
configuration of the experiment, the $^{106}$CdWO$_4$ scintillator
is viewed by a low-background PMT through a PbWO$_4$ light-guide
produced from deeply purified archaeological lead. The new stage
of experiment is now running in the DAMA/Crys low-background
set-up at the Gran Sasso underground laboratory (Italy). The
$^{106}$CdWO$_4$ detector is operated in coincidence with two
large volume CdWO$_4$ crystal scintillators in close geometry
aiming at increase of the detection efficiency to the $2\beta$
processes in $^{106}$Cd with emission of $\gamma$ quanta
\cite{Cerulli:2017}. The sensitivity of the experiment, e.g., to
two neutrino mode of $\varepsilon\beta^+$ decay of $^{106}$Cd is
on the level of the theoretical estimations of the process
half-life $T_{1/2}\sim 10^{22}$ yr.

A large volume cadmium tungstate crystal (mass of the crystal
boule 1868 g) from isotopically enriched cadmium 116
($^{116}$CdWO$_4$) was developed also by the low-thermal-gradient
Czochralski technique \cite{Barabash:2011}. The boule was cut into
several parts, two of them with a total mass 1162 g were used in
the Aurora experiment at the Gran Sasso underground laboratory.
After more than 25 thousand hours of data taking the half-life of
$^{116}$Cd relatively to the two neutrino $2\beta^-$ decay of
$^{116}$Cd is measured with the highest up-to-date accuracy as
$T_{1/2} = [2.69 \pm 0.02(stat.)\pm 0.14(syst.)]\times 10^{19}$
yr. The systematic error of the $T_{1/2}$ value is mainly due to
the radioactive contamination of the crystal by $^{238}$U at the
level of $\sim 0.6$ mBq/kg. A new improved limit on the
$0\nu2\beta^-$ decay of $^{116}$Cd has been set as $T_{1/2} \geq
2.4\times 10^{23}$ yr at 90\% C.L., that corresponds to the limit
on the effective Majorana neutrino mass $\langle m_{\nu} \rangle
\leq (1.2-1.5)$ eV (depending on the nuclear matrix elements used
in the estimations of the neutrino mass limit). New improved
limits on other double beta decay processes in $^{116}$Cd (decays
with majoron emission, transitions to excited levels) were set at
the level of $T_{1/2} > 10^{21}-10^{22}$ yr
\cite{Danevich:2016,Polischuk:2017}. The experiment up to date is
the most sensitive study of the $^{116}$Cd double beta decay.

An important advantage of the low-thermal-gradient Czochralski
method is very high yield of crystalline boules (87\% of the
initial $^{106}$CdWO$_4$ and $^{116}$CdWO$_4$ powders). Besides, a
much lower overheating of the melt (in comparison to the
conventional Czochralski process) allowed to achieve very low
evaporation losses of costly enriched cadmium in the growing
process. The total irrecoverable losses of enriched cadmium in the
$^{106}$CdWO$_4$ and $^{116}$CdWO$_4$ developments does not exceed
2\% (with the main losses appeared due to the purification of the
cadmium samples by vacuum distillation, that can be further
reduced by development of especial equipment to minimize the
losses).

Calcium molybdate crystal scintillators enriched in molybdenum
100, from calcium depleted in calcium 48 were developed for the
AMoRE $2\beta$ experiment \cite{So:2012a,So:2012b} aiming at
search for $0\nu2\beta^-$ decay of $^{100}$Mo by using cryogenic
scintillating bolometers. Using of the depleted calcium is
requested to avoid contribution of the $2\nu2\beta^-$ events of
$^{48}$Ca ($Q_{2\beta}=4268$ keV) to the ROI of the AMoRE
experiment (energy of the expected $0\nu2\beta^-$ peak of
$^{100}$Mo is $Q_{2\beta}=3034$ keV). The crystal scintillators
were produced by the conventional Czochralski method.

Zinc (ZnMoO$_4$) and lithium (Li$_2$MoO$_4$) molybdate crystal
scintillators were extensively studied by the LUMINEU $2\beta$
project to develop high sensitivity detectors to search for
$0\nu2\beta^-$ decay of $^{100}$Mo. High quality crystal
scintillators were obtained after the R\&D that included deep
purification of molybdenum and application of the
low-thermal-gradient Czochralski process to grow crystals
\cite{Chernyak:2013,Berge:2014,Armengaud:2015,Chernyak:2015,Bekker:2016}.
Both the scintillation materials were then produced from enriched
$^{100}$Mo \cite{Barabash:2014,Armengaud:2017}. Similar to the
enriched cadmium tungstate crystals, rather large crystal boules
yield ($\approx 84\%$ and $\approx 84\%$ of the initial charge for
Zn$^{100}$MoO$_4$ and Li$_2^{100}$MoO$_4$, respectively) and low
losses of molybdenum ($\approx4\%$ and $\approx3\%$) were achieved
 in the Zn$^{100}$MoO$_4$ and Li$_2^{100}$MoO$_4$ crystals
production.

The radiopurity level of the developed enriched cadmium tungstate,
lithium, calcium and zinc molybdate crystal scintillators is
presented in Table \ref{tab4}.

\section{Prospects for the next generation $0\nu2\beta^-$ experiments with crystal scintillators}
\label{sec:prosp}

Two big projects aiming at construction of large scale experiments
to search for $0\nu2\beta^-$ decay on the level of inverted
hierarchy of the neutrino masses intend to use the technique of
low temperature scintillating bolometers.

The AMoRE (Advanced Mo-based Rare process Experiment) project
\cite{Bhang:2012,Alenkov:2015} utilizes Metal Magnetic Calorimeter
sensors to measure the heat and scintillation signals from calcium
molybdate crystal scintillators \cite{Lee:2011}. Now the
experiment is running in a pilot phase with $\approx 1.5$ kg of
enriched in $^{100}$Mo and depleted in $^{48}$Ca
$^{48depl}$Ca$^{100}$MoO$_4$ crystal scintillators at the YangYang
underground laboratory (Korea). The next phase of the experiment,
AMoRE I, should start in the beginning of 2018 and utilize the
same low background cryostat with $\approx 5$ kg of
$^{48depl}$Ca$^{100}$MoO$_4$ crystals. A half-life sensitivity of
the experiment over 5 yr of data taking is expected to be at the
level of $T_{1/2}\sim 10^{25}$ years at 90\% C.L. assuming zero
background. A next stage AMoRE II experiment with $\approx200$ kg
molybdate crystal scintillators is planned in a deeper underground
site in Korea that is under construction now. The goal of AMoRE II
is to explore the inverted hierarchy of neutrino mass with a
half-life sensitivity at the level of $T_{1/2}\sim 5\times
10^{26}$ years that corresponds to the effective neutrino mass
$\langle m_{\nu} \rangle \sim 0.02-0.03$ eV (depending on the
nuclear matrix elements calculations). R\&D of several molybdenum
containing crystal scintillators is in progress to replace
CaMoO$_4$ in the AMoRE II phase.

The aim of CUPID (Cuore Upgrade with Particle IDentification) is
to construct a tonne-scale low temperature experiment to probe the
Majorana nature of neutrinos and discover the lepton number
violation in the inverted hierarchy region of the neutrino mass
\cite{CUPID,CUPID-RD}. The experiment is intended to exploit the
current CUORE infrastructure \cite{CUORE-setup} as much as
possible. The isotopes $^{82}$Se, $^{100}$Mo, $^{116}$Cd and
$^{130}$Te, embedded in ZnSe, ZnMoO$_4$ or Li$_2$MoO$_4$, CdWO$_4$
and TeO$_2$ crystals, are considered as the candidate nuclei. A
substantial reduction of background to the level of $<0.02$
counts/(ton$\times$yr) in the ROI can be achieved thanks to
particle identification by simultaneous registration of thermal
and scintillation signals (Cherenkov radiation in the case of
TeO$_2$). With an energy resolution at endpoint (FWHM$~=5$ keV)
CUPID could reach a 90\% C.L. sensitivity of $\lim T_{1/2}\sim
(2-5)\times 10^{27}$ yr after 10 years of operation, that
corresponds to a range of the effective Majorana neutrino masse
sensitivity $\lim \langle m_{\nu} \rangle \sim 6-19$ meV,
depending on the estimate of the nuclear matrix elements.

The CUPID R\&D program is in progress aiming at development of
purification and crystallization procedures, new detector
technologies, development of existing and search for new
scintillation materials containing the candidate elements. A first
array of enriched Zn$^{82}$Se bolometers is running at the Gran
Sasso underground laboratory \cite{Artusa:2016}. A comprehensive
R\&D was performed in the framework of the LUMINEU project to
develop high performance ZnMoO$_4$ and Li$_2$MoO$_4$ cryogenic
detectors for $0\nu2\beta^-$ experiment with $^{100}$Mo. Large
volume (production of scintillation elements 5 cm in diameter and
5 cm height was well established), highly radiopure (activities of
$^{226}$Ra and $^{228}$Th less than 5 $\mu$Bq/kg), high optical
quality crystal scintillators (including two large volume samples
from enriched $^{100}$Mo) were developed \cite{Armengaud:2017}.
Nevertheless, presence of second crystal phase inclusions remains
the main problem both in production and bolometric performance of
ZnMoO$_4$ crystals. In parallel to the ZnMoO$_4$ development, an
R\&D of lithium molybdate was pursued. Several underground tests
of Li$_2$MoO$_4$ and Li$_2^{100}$MoO$_4$ detector modules, as well
as features of the production process, encouraged choice of the
material for the construction of the CUPID-0/Mo demonstrator. The
production of 20 enriched Li$_2^{100}$MoO$_4$ scintillation
elements 4.4 cm in diameter and 4.5 cm height for the first stage
of the experiment was completed recently
\cite{Grigorieva:2017,Shlegel:2017}.

A first test of 34.5 g enriched $^{116}$CdWO$_4$ sample at
milli-Kelvin temperature confirmed a high performance of the
material in terms of energy resolution and particle discrimination
capability \cite{Barabash:2016b}. The aim of the CYGNUS project
(CrYoGenic search for NeUtrinoleSs double beta decay of cadmium)
is investigation of $2\beta^-$ decay of $^{116}$Cd with 1.16 kg of
enriched radiopure $^{116}$CdWO$_4$ crystal scintillators
\cite{Barabash:2011} as cryogenic scintillating bolometers. The
proposed experiment will take advantage of the EDELWEISS ultra-low
background set-up located at the deepest European underground
Laboratory in Modane (France). The project should further prove
suitability of the material for the CUPID project.

Developments of TeO$_2$ detectors with Cherenkov emission
readout\footnote{The idea to use Cherenkov signal for $\beta$ and
$\alpha$ events discrimination was first proposed by Tabarelli de
Fatis in 2010 \cite{Tabarelli:2010}).} are carried out by using
different approaches to detect very weak Cherenkov light from
TeO$_2$ crystals
\cite{Nones:2012,Casali:2015,Battistelli:2015,Schaffner:2015,Artusa:2017}.

\section{Conclusions}
\label{sec:con}

Crystal scintillators are successfully used in $2\beta$
experiments thanks to the possibility to realize calorimetric
"source = detector" approach for several nuclei with a high
detection efficiency. A very low radiopurity level of tungstate
and molybdate crystals (including the scintillators from
isotopically enriched materials) is achieved thanks to the deep
purification of the initial materials, application of the
low-thermal-gradient Czochralski process to grow the crystals, and
2nd crystallization. A high energy resolution (a few keV) and
excellent particle discrimination ability of the low temperature
scintillating bolometers make the detectors promising for the next
generation $0\nu2\beta^-$ experiments aiming at test the inverted
neutrino mass scheme and even go toward the normal neutrino mass
hierarchy.

\section*{Acknowledgments}

Author gratefully acknowledges support from the "Jean d'Alembert"
Grants program (Project CYGNUS) of the University of Paris-Saclay.

\end{document}